\documentclass[twocolumn]{aastex631}

\usepackage{amssymb}
\usepackage{mathrsfs}
\usepackage{amsmath}

\begin{document}

\title{A Phase-resolved View of Millihertz Quasi-periodic Oscillations in the Ultraluminous X-ray Source M51 ULX-7: Evidence for a Magnetically Truncated Disk and Geometrical Beaming}

\correspondingauthor{Shu Zhang}
\email{szhang@ihep.ac.cn}
\correspondingauthor{Hua Feng}
\email{hfeng@ihep.ac.cn}

\author[0000-0001-5160-3344]{Qing-Cang Shui}
\affiliation{Key Laboratory of Particle Astrophysics, Institute of High Energy Physics, Chinese Academy of Sciences, 100049, Beijing, China}
\affiliation{University of Chinese Academy of Sciences, Chinese Academy of Sciences, 100049, Beijing, China}

\author{Shu Zhang}
\affiliation{Key Laboratory of Particle Astrophysics, Institute of High Energy Physics, Chinese Academy of Sciences, 100049, Beijing, China}

\author[0000-0001-7584-6236]{Hua Feng}
\affiliation{Key Laboratory of Particle Astrophysics, Institute of High Energy Physics, Chinese Academy of Sciences, 100049, Beijing, China}

\author[0000-0001-8768-3294]{Yu-Peng Chen}
\affiliation{Key Laboratory of Particle Astrophysics, Institute of High Energy Physics, Chinese Academy of Sciences, 100049, Beijing, China}

\author[0000-0001-5586-1017]{Shuang-Nan Zhang}
\affiliation{Key Laboratory of Particle Astrophysics, Institute of High Energy Physics, Chinese Academy of Sciences, 100049, Beijing, China}
\affiliation{University of Chinese Academy of Sciences, Chinese Academy of Sciences, 100049, Beijing, China}

\author[0000-0002-5554-1088]{Jing-Qiang Peng}
\affiliation{Key Laboratory of Particle Astrophysics, Institute of High Energy Physics, Chinese Academy of Sciences, 100049, Beijing, China}
\affiliation{University of Chinese Academy of Sciences, Chinese Academy of Sciences, 100049, Beijing, China}



\begin{abstract}
X-ray quasi-periodic oscillations (QPOs) are commonly observed in Galactic X-ray binaries (XRBs) and extragalactic ultraluminous X-ray sources (ULXs). In this study, we perform a phase-resolved analysis of recently discovered X-ray millihertz QPOs in M51 ULX-7. This represents the first detailed phase-resolved analysis of QPOs conducted in ULXs. Our findings reveal that the amplitude of the mHz QPO slightly increases with photon energy, accompanied by a narrowing of the phase modulation profile. The phase-resolved spectroscopy indicates significant variability in the energy spectrum: both disk blackbody components exhibit marked variations on the QPO timescale, with the low-temperature component demonstrating significant synchronous changes in the disk temperature and luminosity, showing a positive correlation between these two parameters throughout the QPO cycle. This correlation supports the hypothesis that the disk inner radius corresponds to the magnetospheric radius, which slightly varies with the accretion rate. Our results suggest that the soft component, without beaming, originates from a magnetically truncated outer disk, while the hard component is geometrically beamed from the inner funnel regions.
\end{abstract}



\section{Introduction} \label{sec:intro}
Ultraluminous X-ray sources (ULXs) are off-nuclear, point-like, bright X-ray sources with apparent luminosities exceeding $\sim10^{39}\ {\rm erg\ s^{-1}}$, which result from accretion onto a compact object \citep[see][for comprehensive reviews]{2011NewAR..55..166F,2017ARA&A..55..303K,2023NewAR..9601672K}. 
The remarkable detection of coherent pulsations in M82 ULX-2 \citep{2014Natur.514..202B} demonstrated that ULXs can be powered by magnetic neutron stars (NSs), leading to the classification of these sources as pulsating ULXs (PULXs). To date, among more than 1800 identified ULXs, there are six known extragalactic PULXs, most of which emit at super-Eddington luminosities \citep{2022MNRAS.509.1587W,2024A&A...681A..16T}. 

The discovery of these PULXs marks a significant milestone that poses a challenge to the current understanding of both ULXs and the magnetospheric accretion of NSs \citep[see e.g.][]{2017MNRAS.467.1202M}. PULXs are considered to be driven by NSs with extremely high mass accretion rates and/or exceptionally strong magnetic fields. At such extreme mass accretion rates, geometrically thick, radiation-pressure-dominated accretion disks are expected to produce strong outflows \citep{1973A&A....24..337S,1999AstL...25..508L,2007MNRAS.377.1187P}, which have been observed in both simulations and observations of several sources \citep[see e.g.][]{2011ApJ...736....2O,2014ApJ...796..106J,2016Natur.533...64P,2018MNRAS.473.5680K,2018MNRAS.479.3978K,2020MNRAS.492.4646P}. These outflows may be optically thick and form a funnel-like structure, which plays an important role in shaping emergent X-ray energy spectrum and timing behaviors. For instance, beaming of X-ray emission from the funnel is expected. This beaming affects the apparent luminosity $L_{\rm app}$, enhancing it relative to the accretion luminosity $L$ by a factor of $b^{-1}$. The determination of the beaming factor $b$ remains a subject of debate. Some studies assumed $b$ to be close to unity \citep[see e.g.][]{2015MNRAS.449.2144D,2017MNRAS.467.1202M,2019MNRAS.484..687M,2021MNRAS.501.2424M}, while others proposed a relatively higher beaming factor with $b^{-1}\gtrsim20$ \citep[see e.g.][]{2017MNRAS.468L..59K,2017MNRAS.470L..69M}.

Investigating short-term variability can impose constraints on different accretion models in ULXs \citep[see e.g.][]{2014Natur.513...74P}. Specifically, quasi-periodic oscillations (QPOs) have been detected in the frequency range of $\sim$0.5--600 mHz in a few ULXs \citep[e.g. M82 X-1, Holmberg IX X-1, NGC 5408 X-1, NGC 6946 X-1, M82 X-2, IC 342 X-1, 4XMM J140314.2+541806 and M51 ULX-7,][]{2003ApJ...586L..61S,2006ApJ...641L.125D,2007ApJ...660..580S,2010ApJ...722..620R,2010ApJ...710L.137F,2015MNRAS.446.3926A,2022MNRAS.511.4528U,2024A&A...689A.284I}. The origin of QPOs in XRBs, and especially in ULXs, remains a topic of debate, with several alternative explanations proposed \citep[see][for a review]{2019NewAR..8501524I}. The millihertz frequency range is too low to align with Keplerian rotation in the region responsible for X-ray emission. Instead, these frequencies might be associated with the Lense-Thirring precession of inflows/outflows \citep{2018MNRAS.475..154M,2019MNRAS.489..282M}. Additionally, another form of quasi-periodic variability, observed at frequencies around 1 mHz in the ULX 4XMM J111816.0–324910 \citep{2020ApJ...898..174M}, which resembles the so-called ``heartbeat" variability typically seen in the Galactic black hole XRB GRS 1915+105 \citep{2000A&A...355..271B}. This ``heartbeat" variability has often been interpreted as limit-cycle instabilities within the inner accretion disk, likely arising from radiation pressure instability \citep{1974ApJ...187L...1L,1998MNRAS.298..888S,2000ApJ...542L..33J,2000ApJ...535..798N}.

M51 consists of an interacting galaxy pair, which includes the active, face-on spiral galaxy NGC 5194 and its companion, the dwarf galaxy NGC 5195. The X-ray source known as M51 ULX-7 (hereafter referred to as ULX-7) was first identified by the Einstein X-ray Observatory in these galaxies \citep{1985ApJ...298..259P}. Positioned in a spiral arm to the northwest of the center of NGC 5194, ULX-7 exhibits significant variability. A recent study by \citet{2020ApJ...895...60R} determined a spin period of $P_{\rm spin}\approx2.8$ s, and that the neutron star was in a 2 day orbit with a $>8 M_{\odot}$ companion star, categorizing it as a PULX within a high-mass XRB system. Assuming that variations in the mass accretion rate cause the superorbital modulation, the measured spin-up rate $\dot{P}_{\rm spin} \approx -2.4 \times 10^{-10}\ {\rm s\ s^{-1}}$ indicates a moderately strong magnetic field with $B=10^{12}-10^{13}$ G \citep{2020ApJ...895...60R}. Furthermore, \citet{2021ApJ...909....5H} and \citet{2021ApJ...909...50V} also found evidence for periodic dips in the \emph{Chandra} X-ray light curve that are associated with the 2 day binary orbital period, which suggests an inclination angle of $i\sim60^\circ$. Recently, \citet{2024A&A...689A.284I} detected QPOs at frequencies of $\sim0.5$ mHz in ULX-7 using three \emph{XMM-Newton} observations in 2021 and 2022. In this study, we conducted a novel phase-resolved analysis of the newly discovered mHz QPOs in ULX-7. We provide an overview of the observations and our data reduction in Section~\ref{sec:2}, followed by the presentation of the timing and spectral analyses in Section~\ref{sec:3}. Finally, we discuss and summarize these results in Section~\ref{sec:4}.

\section{Observations and Data Reduction} \label{sec:2}
In this study, we performed a phase-resolved analysis of the mHz QPOs in ULX-7 using data from three \emph{XMM-Newton} observations (ObsIDs 0883550101, 0883550201, and 0883550301). For convenience, we will refer to these observations as observations 101, 201, and 301, respectively. Details of the \emph{XMM-Newton} observations utilized in this study can be found in Table~\ref{tab:1}.

The data were processed using the \emph{XMM-Newton} \textsc{Science Analysis System} (\textsc{xmmsas}) version 19.1.0\footnote{\url{https://www.cosmos.esa.int/web/XMM-Newton}} with the latest calibration files (February 2021). The raw data were obtained from the XMM-Newton Science Archive (XSA)\footnote{\url{https://www.cosmos.esa.int/web/XMM-Newton/xsa}}. We executed the \textsc{epproc} and \textsc{emproc} tasks to generate the EPIC-PN and EPIC-MOS event files, respectively.  Following the methodology outlined by \citet{2024A&A...689A.284I}, events were extracted from a circular region with a radius of $20''$ centered on the source position \citep[RA = 13h30m01s.02, Dec = 47$^\circ13'43''.8$, J2000;][]{2016ApJ...827...46K} and were selected to have \texttt{PATTERN $\le4$} for the EPIC-PN data and \texttt{PATTERN $\le12$} for the EPIC-MOS data. The background was estimated using an annular region centered on the source position, with inner and outer radii of $21''$ and $39''$, respectively. To identify periods of high background, we created light curves of the events in the 10--15 keV band. For the timing analysis, we removed only the particle flares at the beginning and/or end of the observation to minimize additional gaps in the light curve. The times of arrival (ToAs) of the photons were corrected to the barycenter of the solar system using the \textsc{xmmsas} task \textsc{barycen}. Subsequently, the background-corrected light curves for ULX-7 were generated using the \textsc{epiclccorr} task. We confirmed that the background flares had no imprint on the background-subtracted light curves, and the source to background flux ratio in the source extraction region during high-background intervals was always higher than 15. Additionally, we examined the timing gaps resulting from that the total count rate exceeded the EPIC-PN and -MOS telemetry limits\footnote{\url{https://heasarc.gsfc.nasa.gov/docs/xmm/uhb/epicmode.html}} ($\sim 600$ and $\sim115$ counts $\rm s^{-1}$, respectively) and found no gaps longer than the time bin size (100 s) used in the timing analysis. For the spectral analysis, we further excluded high background intervals that occurred during the observations. Response matrices and ancillary response files were created using the \textsc{xmmsas} tasks \textsc{rmfgen} and \textsc{arfgen}. Finally, the source spectra were grouped using the tool \textsc{specgroup} to a minimum of 25 counts per spectral bin and at the same time not to oversample the instrument energy resolution by more than a factor of 3\footnote{\url{https://heasarc.gsfc.nasa.gov/docs/xmm/sas/help/specgroup/node15.html}}. This allows the application of $\chi^2$ statistics in the spectral modeling.

\begin{table}[]
    \centering
    \caption{Log of \emph{XMM-Newton} Observations of M51 Used in This Work.\label{tab:1}}
    \begin{tabular}{cccc}
    \hline
    \hline
    \# & Observation ID & Start Time & Exposure Time$^{(\rm a)}$\\ 
       &                & (MJD) & (ks)\\
    \hline
    1 & 0883550101 & 59540.40 & 130.4 \\
    2 & 0883550201 & 59542.39 & 130.2 \\
    3 & 0883550301 & 59586.25 & 131.4 \\
    \hline
    \hline
    \end{tabular}
    \tablecomments{$^{(\rm a)}$ Pre-flare filtering exposure time.}
\end{table}

\begin{figure*}
\centering
    \includegraphics[width=\linewidth]{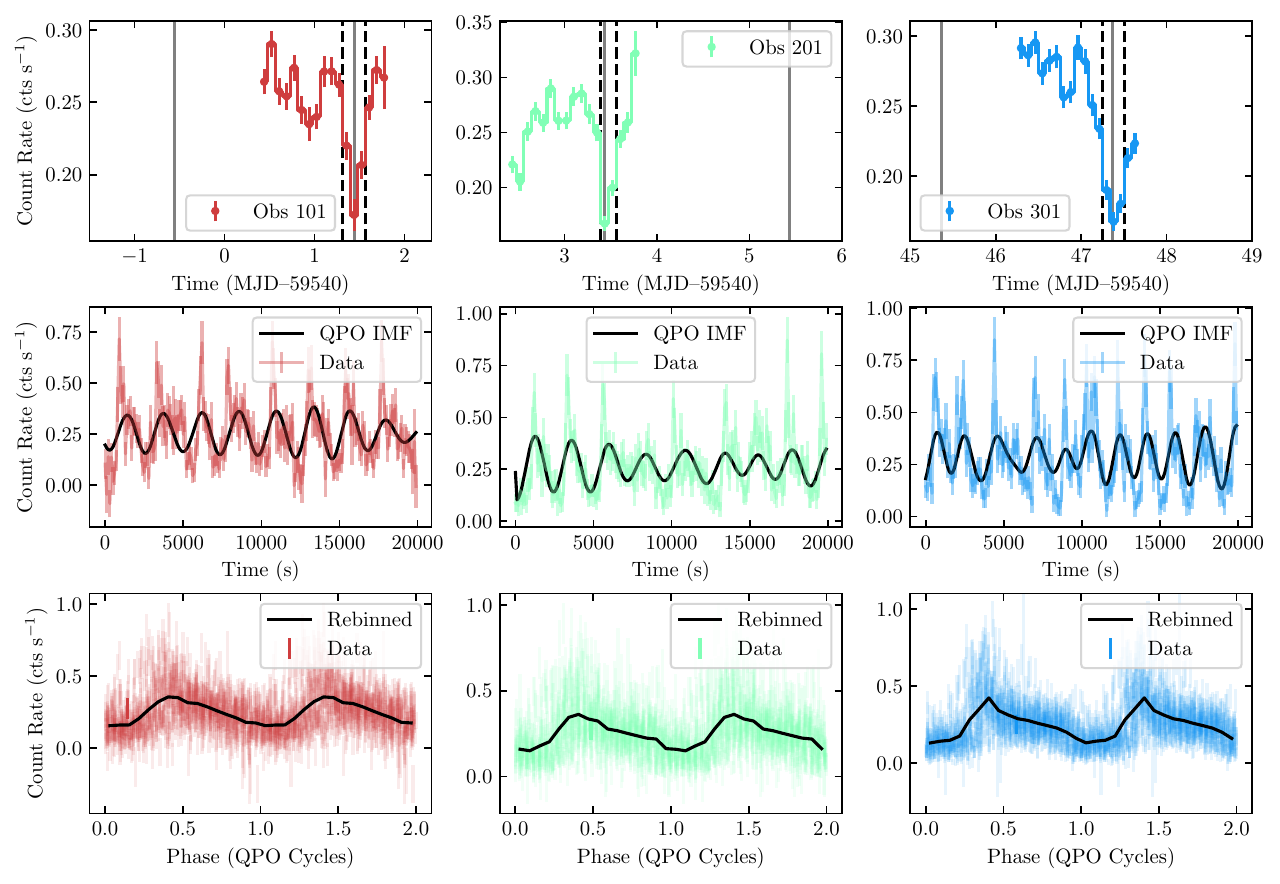}
    \caption{Results of the timing analysis. \emph{Top}: light curves of M51 ULX-7 with a 2-hour time resolution from the three observations 101, 201 and 301. Black dashed lines indicate the edges of dips obtained from the Bayesian block technique and gray solid lines indicate the expected arrival times of the dips, assuming the dips occur periodically with an orbital period of 1.9969 days, using the dip time from observation 101 as the reference point. \emph{Middle}: representative 20 ks light curves (colored points with error bars) with a 100-s time bin from three observations, and the intrinsic QPO light curves (black solid lines) obtained from HHT analysis. \emph{Bottom}: count rates are plotted against QPO phase as colored points with error bars and HHT phase-folded light curves are plotted with 20 phase bins per cycle as black solid lines.}
    \label{fig:1}
\end{figure*}

\section{Analysis and Results}
\label{sec:3}
\subsection{Timing Analysis}
We first performed a timing analysis for the three observations. The top panels of Figure~\ref{fig:1} present PN+MOS background-subtracted light curves in the 0.3--10 keV energy range, with a time resolution of 2 hours. This resolution was selected to optimize the detection of long-term ($\gtrsim$ several hours) flux variability within individual observations, specifically to probe the periodic flux dips reported in previous studies of M51 ULX-7 \citep{2021ApJ...909....5H,2021ApJ...909...50V}. The 2-hour resolution provides a balance between acceptable statistics and sufficient timing resolution to resolve variability on timescales comparable to the dip durations observed previously ($\sim8$ hours). Three dips are clearly visible in these observations; the dips in observations 101 and 201 are fully observed, whereas the dip in observation 301 appears to be partially seen. To quantitatively characterize the dip features and assess their significance, we employed the Bayesian block\footnote{\url{https://docs.astropy.org/en/stable/api/astropy.stats.bayesian_blocks.html}} technique to identify these dips \citep{2013ApJ...764..167S}. With a false alarm probability of $p_0=0.005$, we identified four significant flux variations. Three of these variations correspond to visible dips, the edges of which are indicated by black dashed lines in the top panels of Figure~\ref{fig:1}. This suggests that the flux during these dip periods significantly deviates from the long-term flux trend. The central times of the three dips were calculated from the center of the blocks as MJD 59541.44, 59543.48, and 59587.38, respectively. In the top panels of Figure \ref{fig:1}, the expected arrival times of the dips are plotted as gray solid lines, assuming the dips occur periodically with an orbital period of 1.9969 days \citep{2020ApJ...895...60R}, using the dip time from observation 101 as the reference point. It is evident that the observed dip features occur periodically with each binary orbit. The estimated widths of these dips are $0.23\pm0.07$, $0.22\pm0.08$, and $0.32\pm0.09$ days, consistent with dip features observed by \emph{Chandra} in 2012 \citep{2021ApJ...909....5H,2021ApJ...909...50V}. The errors correspond to the standard deviation obtained from bootstrapping. Specifically, bootstrapping involves randomly selecting (with replacement) $N$ times from $N$ bins of data. For our light curves, which consist of 51 time bins, we found that 788 bootstrap iterations were necessary to achieve convergence on the true error distribution\footnote{following the $N\cdot(\ln{N})^2$ criterion; see \citet{2012msma.book.....F}}. In each bootstrap iteration, we determined the dip widths using the Bayesian blocks technique. The uncertainties were then estimated as the standard deviations of the width distributions obtained from all bootstrap iterations.

The middle panels of Figure~\ref{fig:1} display representative PN+MOS background-subtracted light curves, each covering a period of 20 ks in the 0.3--10 keV energy range, with a time resolution of 100 s. The three light curves exhibit strong, highly structured variability with large amplitudes. The power density spectrum (PDS) can be effectively modeled with several components. One of these components is a Lorentzian function centered at frequencies of $f \sim 0.5$ mHz, with a quality factor $Q = f/\Delta f \sim 2-10$, where $\Delta f$ is the full width at half maximum (FWHM) of the Lorentzian function (see Appendix~\ref{appendix1}). Consequently, this PDS component is classified as a mHz QPO \citep[see][for details]{2024A&A...689A.284I}. In addition to the mHz QPO, the PDS also exhibits other continuum components, including a broader Lorentzian feature ($Q<2$) at higher frequencies and a power-law component that dominates at frequencies of $f<10^{-5}$ Hz. However, in this study, we focus solely on the phase-resolved analysis of the mHz QPO component. Due to the short timescale ($\sim10$ ks) variations in the recurrence time and amplitude of this QPO variability, period folding for phase-resolved analysis is not suitable. Therefore, we utilized the Hilbert-Huang Transform (HHT) analysis procedure described by \citet{2023ApJ...957...84S} to conduct the phase-resolved analysis of the mHz QPO. The HHT method, originally introduced by \citet{1998RSPSA.454..903H} as an adaptive data analysis technique, serves as a powerful tool for studying signals with non-stationary periodicity \citep{1998RSPSA.454..903H,2008RvGeo..46.2006H}. This method consists of two main components: mode decomposition and Hilbert spectral analysis (HSA). Mode decomposition can decompose a time series into several intrinsic mode functions (IMFs), while HSA enables the extraction of both the phase function and instantaneous frequency for the desired IMFs, such as QPOs \citep[see e.g.][]{2014ApJ...788...31H,2020ApJ...900..116H,2023ApJ...951..130Y,2024ApJ...961L..42Z,2024ApJ...973...59S,2024A&A...692A.117D} and gravitational wave signals \citep[see e.g.][]{2022ApJ...935..127H,2024PhRvD.110j4020S}. We applied the HHT analysis to the three 100-s time bin PN+MOS background-subtracted light curves within the 0.3--10 keV energy range. By employing variational mode decomposition \citep{2014ITSP...62..531D}, we were able to extract the intrinsic QPO light curves, which are illustrated as solid black lines in the middle panels of Figure \ref{fig:1}. Through the application of HSA, we derived the instantaneous phase function of the mHz QPO from each observation. Using these phase functions, the bottom panels of Figure~\ref{fig:1} plot the count rate as a function of phase. By re-binning data points into 20 phase bins per cycle, the phase-folded light curves are depicted as solid black lines in the bottom panels of Figure~\ref{fig:1}. These phase-folded light curves align well with the oscillation features observed in the original light curves and display a nonsinusoidal shape, marked by a rapid ascent followed by a gradual decline.

\begin{figure*}
\centering
\begin{minipage}[c]{0.35\linewidth}
\centering
    \includegraphics[width=\linewidth]{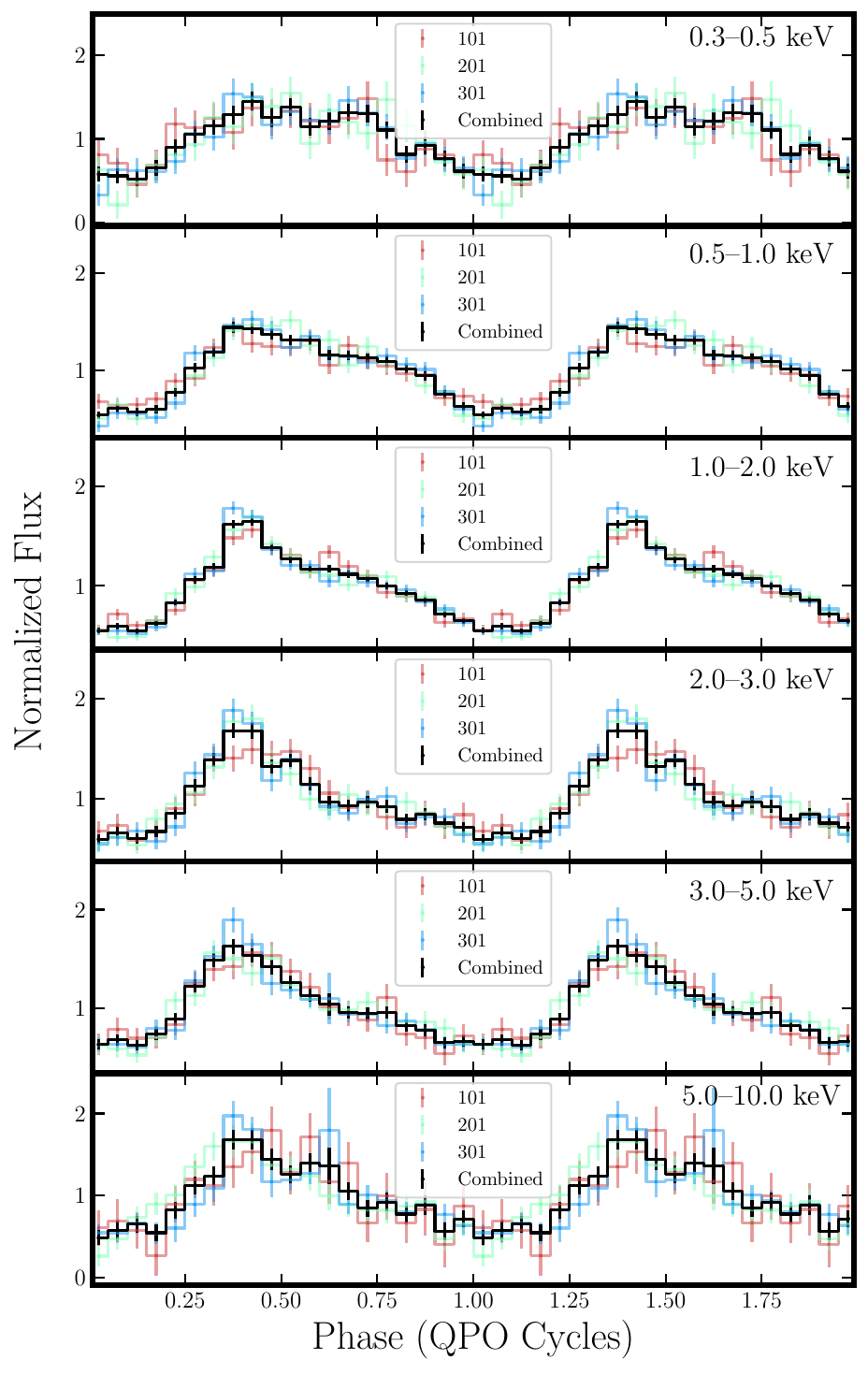}
\end{minipage}
\begin{minipage}[c]{0.35\linewidth}
\centering
    \includegraphics[width=\linewidth]{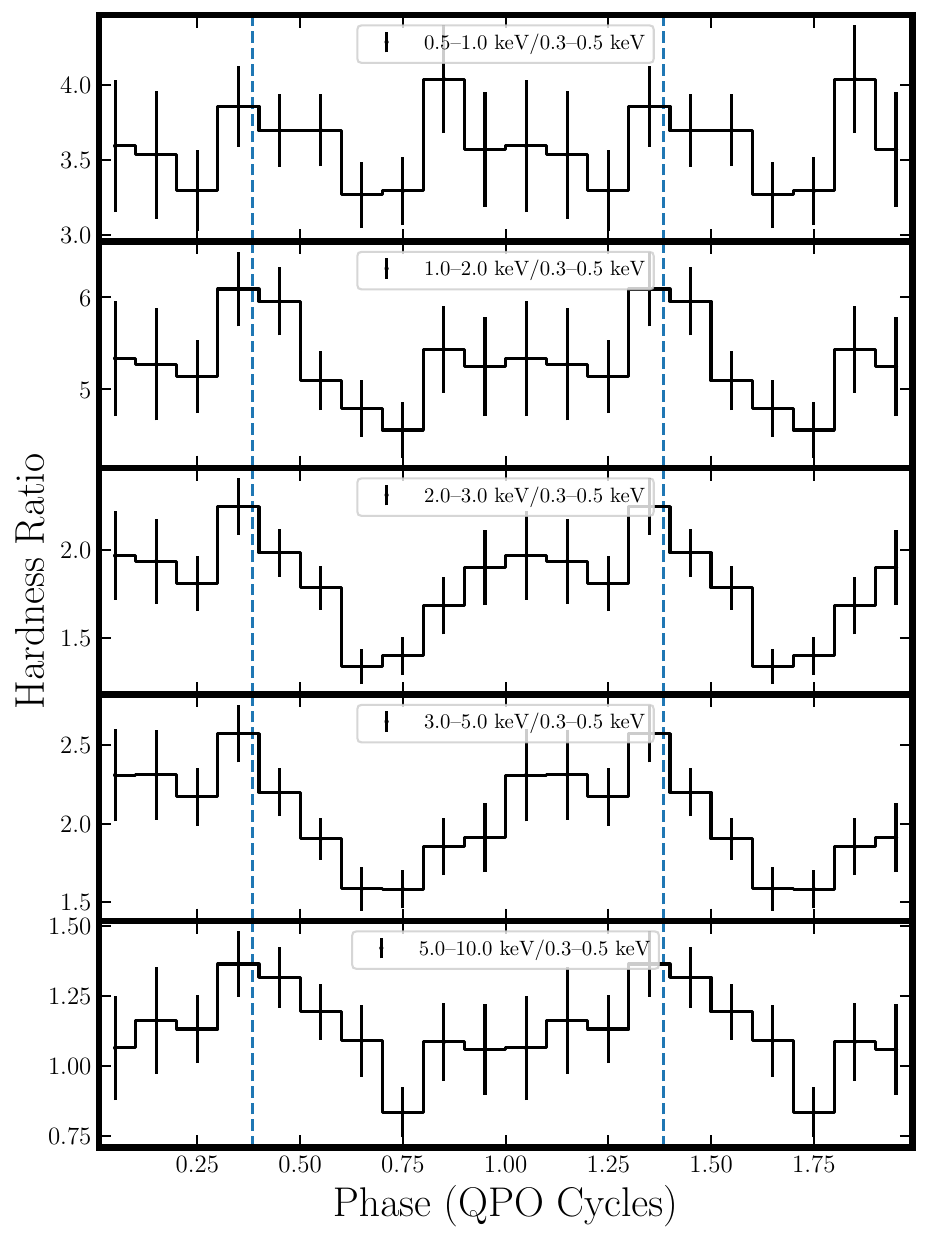}
\end{minipage}
    \caption{\emph{Left}: phase-folded light curves for various energy ranges from observations 101 (red), 201 (green), 301 (blue), and the combined data from all three observations (black). \emph{Right}: hardness ratios plotted as functions of QPO phase obtained from the combined data. The higher energy bands are defined as follows, from top to bottom: 0.5--1 keV, 1--2 keV, 2--3 keV, 3--5 keV, and 5--10 keV, while the lower energy band is consistently set at 0.3--0.5 keV. The blue dashed lines indicate the peak phases of the folded light curve in the energy range of 0.3--10 keV.} \label{fig:2}
\end{figure*} 

To study the energy dependence of the mHz QPO, we performed phase folding on light curves across various energy ranges. The left panel of Figure~\ref{fig:2} displays the phase-folded light curves for various energy ranges derived from three observations. It is evident that the QPO folding profile are stable and consistent across the three observations. Therefore, to improve the counting statistics of the folded light curves, we combined the data from the three observations, which are illustrated in black. The modulation of the low-energy ($\lesssim1$ keV) folded light curves maintain a relatively symmetrical appearance, exhibiting an overall morphology that resembles a sinusoidal shape. In contrast, the phase-folded light curves in the higher energy ranges ($\gtrsim1$ keV) show a pronounced nonsinusoidal shape, with a rapid rise in the phases of $\sim0-0.4$ cycles and a slower decay in the phases of $\sim0.4-1$ cycles. Overall, the flare (QPO) profiles at lower energies are broader compared to those observed at higher energies. The right panel of Figure~\ref{fig:2} plots the hardness ratios as functions of the QPO phase obtained from the combined folded light curves, using 10 phase bins per cycle. The dashed blue lines mark the peak phases of the folded light curve within the energy range of 0.3--10 keV. Significantly, the hardness ratios seem to attain their maximum values at the phases corresponding to the flux peak. 

\begin{figure}
\centering
    \includegraphics[width=\linewidth]{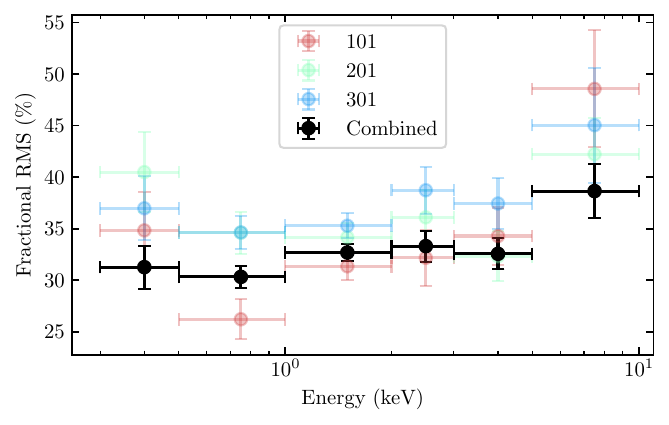}
    \caption{Energy dependence of the fractional RMS ($f_{\rm rms}$) of the mHz QPO from observations 101 (red), 201 (green), 301 (blue), and the combined data from all three observations (black). $f_{\rm rms}$ is computed from Eq.~(\ref{equation1}).} \label{fig:3}
\end{figure}

The fractional root mean square (RMS) for the mHz QPO waveform, denoted as $f_{\rm rms}$, was computed for each energy range shown in the left panel of Figure~\ref{fig:2}. The definition of $f_{\rm rms}$ is given by the following equation:
\begin{equation}\label{equation1}
    f_{\rm rms}=\frac{\left[\sum_{i=1}^N(r_i-\bar{r})^2/N\right]^{1/2}}{\bar{r}}
\end{equation}
where $\bar{r}$ is the phase averaged count rate, $r_i$ is the count rate at the $i$-th phase bin, and $N=20$ is the total phase bin number. The energy dependence of $f_{\rm rms}$ is plotted in Figure~\ref{fig:3}. Despite some slight variations in $f_{\rm rms}$ across different observations, it is evident that the amplitude of the mHz QPO shows an increasing trend over the 0.3--10 keV energy range in each observation. Notably, the rms spectra from observations 101 and 201 exhibit a tentative U-shaped profile, though these trends are not statistically significant.

\subsection{Energy Spectral Analysis}
\begin{figure*}
\centering
\begin{minipage}[c]{0.45\linewidth}
\centering
    \includegraphics[width=\linewidth]{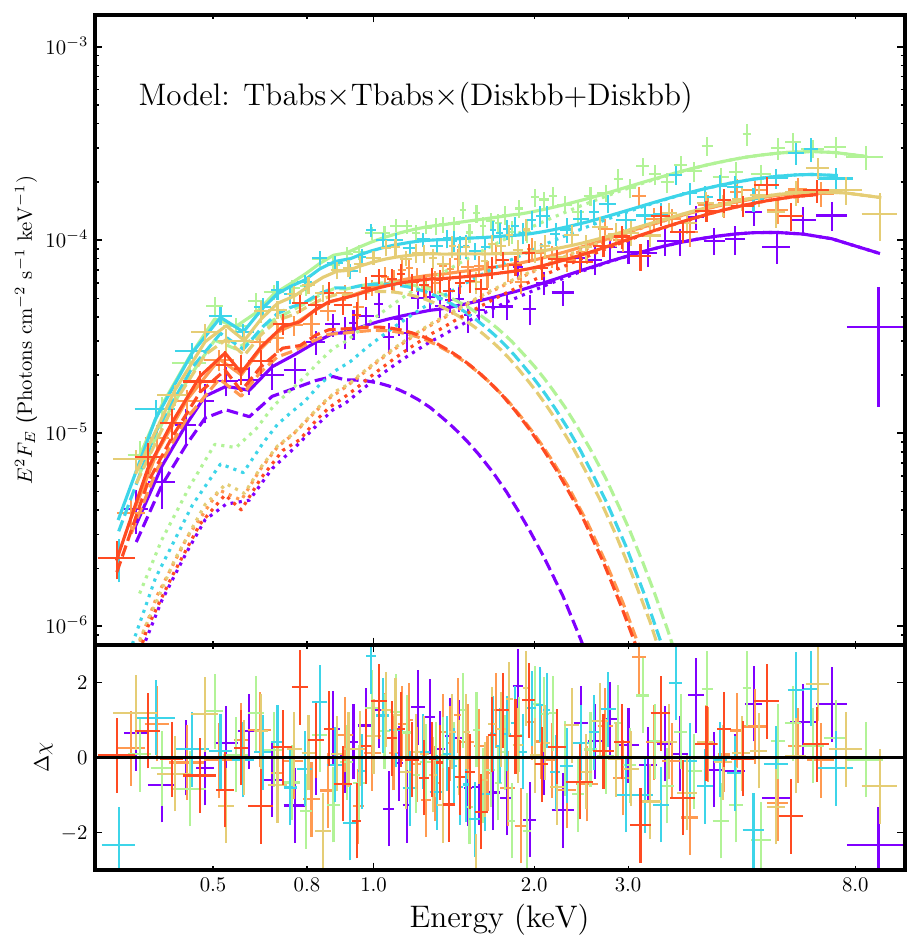}
\end{minipage}
\begin{minipage}[c]{0.48\linewidth}
\centering
    \includegraphics[width=\linewidth]{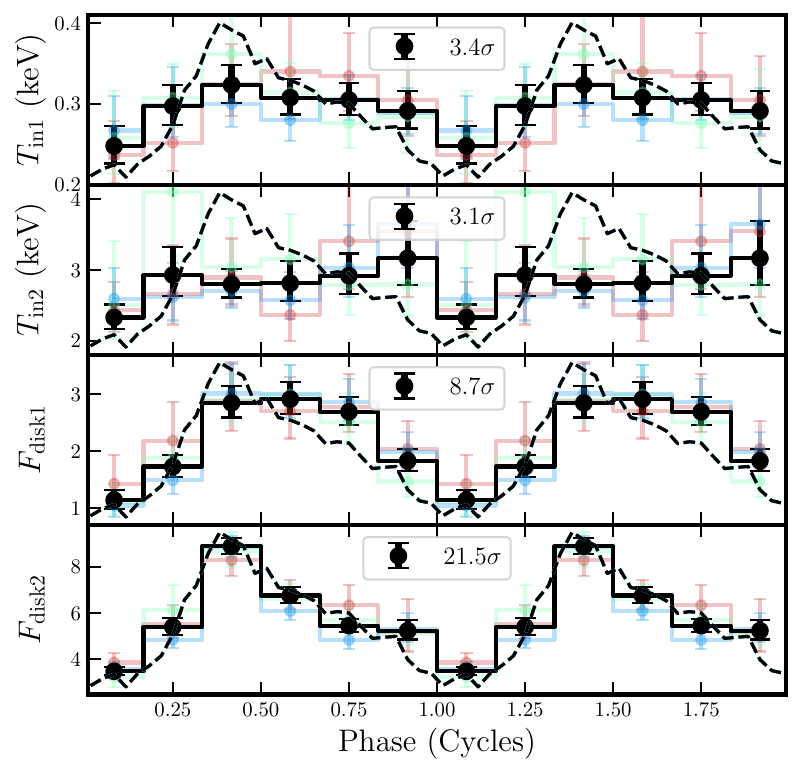}
\end{minipage}
    \caption{Phase-resolved spectral analyses using model: \textsc{tbabs}$\times$\textsc{tbabs}$\times$(\textsc{diskbb}+\textsc{diskbb}). \emph{Left}: jointly spectral fittings of the phase-resolved spectra (top panel) and corresponding residual plots (bottom panel) from the combined data of the three observations. \emph{Right}: variability of spectral parameters during the mHz QPO cycle obtained from the spectral analyses of observations 101 (red), 201 (green), 301 (blue) and combined data from three observations (black). The shown model fluxes ($F_{\rm disk1}$ and $F_{\rm disk2}$) are in units of $\rm 10^{-13}\ erg\ s^{-1}\ cm^{-2}$. Additionally, the phase-folded light curve is included in each panel as a black dashed line. The statistical significance of each modulation from the combined data is quoted in the corresponding plot.} \label{fig:4}
\end{figure*}

Using the well-defined phase functions obtained from the HHT analysis, we extracted EPIC-PN spectra from six distinct phase bins for the three observations, allowing for a subsequent phase-resolved spectral analysis. We conducted the phase-resolved spectral analysis for each observation individually, as well as for the combined data from the three observations, which provided improved counting statistics. The spectral analysis was performed using the \textsc{xspec v12.14.1} software package \citep{1996ASPC..101...17A}. Following \citet{2022ApJ...925...18B} and \cite{2024A&A...689A.284I}, we modeled the spectra with two absorbed multi-temperature disk black bodies: \textsc{tbabs}$\times$\textsc{tbabs}$\times$(\textsc{diskbb}+\textsc{diskbb}). \textsc{tbabs} model was used to model the interstellar absorption \citep{2000ApJ...542..914W}, with a Milky Way component fixed at $3.3\times10^{20} \rm{cm^{-2}}$ and a free extragalactic component. To prevent degeneracy in the spectral parameters, we linked the absorption column density of the extragalactic component across the spectra at different phases. By jointly fitting the six phase-resolved spectra with the aforementioned model, we obtained reasonable $\chi^2/{\rm d.o.f}$ values of 234.82/268, 305.64/293, 360.76/347 and 266.15/268 for observations 101, 201, 301 and combined data from the three observations, respectively. The left panel of Figure~\ref{fig:4} illustrates the joint spectral fitting for the combined data, clearly showing that there are no strong features in the residual plots. Additionally, we employed the \textsc{cflux} model to calculate the unabsorbed flux contributions from the two \textsc{diskbb} components. The uncertainties reported for the fitting model parameters represent a 68\% confidence range ($1\sigma$).

The right panel of Figure~\ref{fig:4} illustrates the variability of spectral parameters throughout the mHz QPO cycle, derived from the joint fitting of the phase-resolved spectra from observations 101 (red), 201 (green), 301 (blue), and the combined data from all three observations (black) using the aforementioned spectral model. Additionally, the phase-folded light curve of the combined data is included in each panel as a black dashed line. We computed the difference between the maximum and minimum values of each parameter within a cycle to assess the significance of parameter modulation. The results clearly demonstrate that parameters from both \textsc{diskbb} components exhibit significant variations during the QPO cycle ($>3\sigma$). Notably, the low-temperature component shows significant synchronous changes in disk temperature and flux. The flux modulation of the low-temperature \textsc{diskbb} component ($F_{\rm disk1}$) is relatively symmetrical, resembling a sinusoidal shape in its overall morphology. In contrast, the flux of the high-temperature component ($F_{\rm disk2}$) exhibits a narrower and more distinctly nonsinusoidal profile. The differing characteristics of the fluxes from the two components align well with the findings presented in the left panel of Figure \ref{fig:2}, where the phase-folded light curves across different energy ranges display distinct profiles.

\begin{figure}
\centering
    \includegraphics[width=\linewidth]{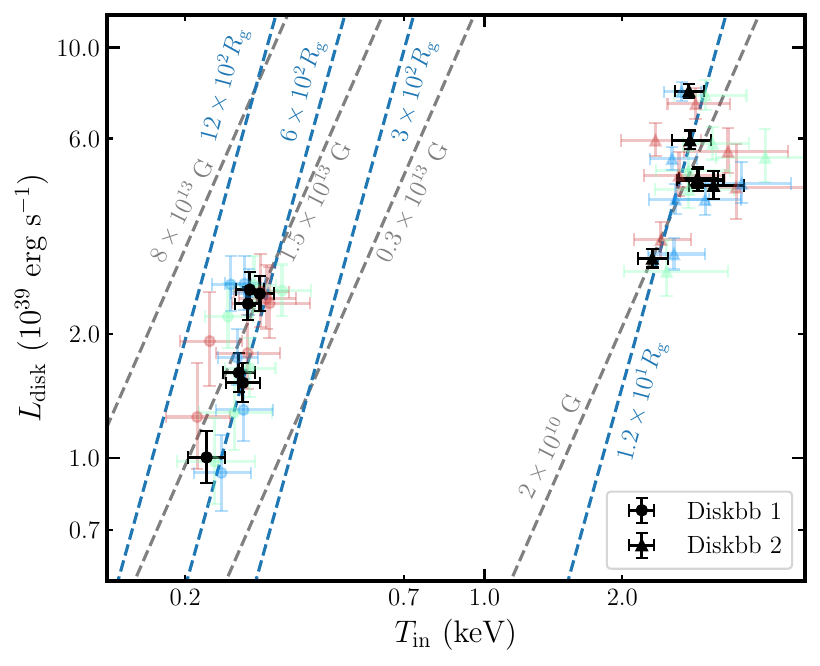}
    \caption{The disk luminosity $L_{\rm disk}$ vs. $T_{\rm in}$ over the mHz-QPO cycle for the low-temperature (circles) and high-temperature (triangles) components from jointly fittings of observations 101 (red), 201 (green), 301 (blue) and combined data from the three observations (black). The dashed blue lines are plotted assuming $L_{\rm disk}= 4\pi\sigma R_{\rm in}^2 T_{\rm in}^4$ , where $R_{\rm in}$ remains constant over the QPO cycle. The dashed gray lines are plotted assuming $R_{\rm in}=R_{\rm m}$ over the QPO cycle.} \label{fig:5}
\end{figure}

As mentioned above, both of two disk components exhibit strong variability throughout the QPO cycle. In Figure \ref{fig:5}, we illustrate the behavior of these two disk components by plotting their luminosities against their temperatures. To derive the unabsorbed component luminosities, we considered a distance from the source of 8.58 Mpc \citep{2016ApJ...826...21M}. For a standard thin disk, the luminosity can be approximated as $L_{\rm disk}\approx4\pi R^2_{\rm in}\sigma T^4_{\rm in}$, where $R_{\rm in}$, $T_{\rm in}$ and $\sigma$ are the inner radius, the temperature and the Stefan-Boltzmann constant, respectively. In the case of PULXs, the accretion disk is interrupted at the magnetospheric radius $R_{\rm m}$, where the magnetic pressure becomes comparable to the ram pressure of the accreting gas. Inside this radius, the accreting plasma is funnelled along the magnetic field lines onto the magnetic poles of the central compact object. The magnetospheric radius can be calculated using the formula:
\begin{equation}
R_{\rm m}\approx7\times10^7\Lambda B_{12}^{4/7}R^{10/7}_6M^{1/7}L_{39}^{-2/7} {\rm cm},
\end{equation}
where $\Lambda$ is a parameter that takes into account the geometry of the accretion flow and in the case of an accretion disk is $\approx0.5$, $B_{12}$ is magnetic dipolar field strength in units of $10^{12}$ G, $R_6$ is the neutron star radius in units of $10^6$ cm, $M$ is the neutron star mass in units of $M_{\odot}$, and $L_{39}$ is luminosity in units of $10^{39}$ erg s$^{-1}$. Assuming $R_{\rm in}=R_{\rm m}$ and $L=L_{\rm disk}$, the relationship between the disk luminosity and temperature can be expressed as
\begin{equation}
L_{\rm disk}=5.8\times10^{39}M^{2/11}R^{20/11}_6B_{12}^{8/11}T^{28/11}_{\rm in} {\rm erg\ s^{-1}},
\end{equation}
where $T_{\rm in}$ is in units of keV, and $\Lambda$ is assumed to be $\approx0.5$. In Figure \ref{fig:5}, we plot the theoretical $L_{\rm disk}\propto T_{\rm in}^4$ and $L_{\rm disk}\propto T_{\rm in}^{28/11}$ relationships as dashed blue and gray lines, respectively, assuming different values for $R_{\rm in}$ and magnetic dipole field strengths. Regarding the low-temperature disk component, within the error range, both theoretical relations can conform to the data points. If we assume that the inner radius does not change during the QPO cycle, it can be approximated as $\sim600R_{\rm g}\approx1.2\times10^8$ cm, where $M$ is assumed to be $1.4M_{\odot}$. Taking into account $R_{\rm in} = R_{\rm m}$, the magnetic field strength can be estimated at $\sim1.5\times10^{13}$ G, which is consistent with previous estimates \citep[$\sim10^{12}-10^{13}$ G,][]{2020ApJ...895...60R,2021ApJ...909...50V}. In contrast, the data points for the high-temperature disk component show significant deviations from the two theoretical relationships. 

\section{Discussion and conclusion} \label{sec:4}
This study has conducted a comprehensive phase-resolved analysis of $\sim0.5$ mHz QPOs in M51 ULX-7. Previous theoretical studies have proposed that mHz QPOs, and day time-scale periods/QPOs (e.g. superobital modulation) are associated with Lense-Thirring precession of the inflow and outflowing wind, respectively \citep{2018MNRAS.475..154M,2019MNRAS.489..282M}. In this Lense-Thirring scenario, variability of X-ray mHz QPOs arises from the geometric wobble of the accretion disk, which alters the projection area of the disk relative to the observer, thereby modulating the observed X-ray flux \citep{2009MNRAS.397L.101I}. This leads to the expectation that the temperature of the disk would not be modulated in sync with the QPO cycle. Contrary to this expectation, our phase-resolved analysis revealed significant variations in spectral shape, particularly demonstrating synchronous variations in disk temperature and flux. Furthermore, if the Lense-Thirring scenario is indeed applicable, precession should primarily occur within the spherization radius $R_{\rm sph}$ \citep{2018MNRAS.475..154M}, where the radiation force from the energy released in the disk is no longer balanced by gravity, causing the disk to become geometrically thick and launch powerful outflows \citep{1973A&A....24..337S}. Under this assumption, if the low-temperature component originates from a region outside of $R_{\rm sph}$, it would not be expected to exhibit modulation with the QPO phase. However, our findings indicate that both disk components display significant modulations throughout the QPO cycle. Thus, we conclude that existing models of the Lense-Thirring scenario do not adequately account for our results, suggesting the need for further exploration and refinement of theoretical frameworks to explain the observed phenomena in this system.

An alternative possibility is that the mHz QPOs arise from limit-cycle instabilities within the accretion disk, driven by the radiation pressure instability \citep{1974ApJ...187L...1L,1998MNRAS.298..888S,2000ApJ...542L..33J}. This mechanism has been extensively considered as the potential origin of the class $\rho$ ``heartbeat'' variability observed in GRS 1915+105 \citep{2000A&A...355..271B}. We found that the mHz-QPO waveform in the 0.3--10 keV energy range exhibits a nonsinusoidal feature, characterized by a fast rise and slow decay.  This pattern contrasts with the class $\rho$ ``heartbeat'' variability, which displays a slow rise followed by a rapid decay \citep{2011ApJ...737...69N}. Furthermore, the class $\rho$ ``heartbeat'' variability is characterized by a more stable variability pattern than the one seen in the ULX-7 light curves. Notably, the flux modulation shape we observed in ULX-7 is similar to that of Class X ``heartbeat''-like variability seen in IGR J17091--3624 \citep{2024ApJ...963...14W,2024ApJ...973...92S}. Additionally, both variability patterns exhibit synchronous variations in the temperature and flux of the low-temperature disk component. Therefore, the observed mHz QPOs, akin to Class X ``heartbeat"-like variability in IGR J17091--3624, could originate from instabilities within the accretion disk, resulting in quasi-periodic fluctuations in the accretion rate. 

By plotting $L_{\rm disk}$ against $T_{\rm in}$ throughout the QPO cycle for both disk components and comparing the data points with theoretical relations, we demonstrated that the low-temperature component in the spectrum from M51 ULX-7 originates from an accretion disk truncated at the magnetospheric radius $R_{\rm m}$, rather than from outflowing winds. This conclusion is supported by the fact that if the low-temperature component was associated with outflowing winds, one would not expect a positive correlation between luminosity and temperature \citep{2009MNRAS.398.1450K,2016MNRAS.456.1837S,2023NewAR..9601672K}. Assuming that the inner radius of low-temperature disk is equal to $R_{\rm m}$, the inferred magnetic field strength is $\sim1.5 \times 10^{13}$ G, which is consistent with previous estimates \citep[$\sim10^{12}-10^{13}$ G,][]{2020ApJ...895...60R,2021ApJ...909...50V}.

Since the low-temperature disk component originates from the magnetically truncated outer disk, its luminosity variability roughly reflects fluctuations in the accretion rate, particularly if the X-ray mHz QPO variability is driven by the accretion rate variations. Consequently, the modulation of the accretion rate throughout the QPO phase mirrors that of the low-energy flux, resembling a roughly sinusoidal shape (see Figure~\ref{fig:2}). The high-temperature component may originate from the accretion curtain \citep{2017MNRAS.467.1202M} and/or the inner thick accretion disk ($R<R_{\rm sph}$). The accretion curtain is a magnetically confined, optically thick structure that channels material from the inner accretion disk to the magnetic poles of the central compact object. In these regions, strong optically thick outflowing winds are effectively launched from the accretion flow, forming a funnel-like structure \citep[see e.g.][]{2007MNRAS.377.1187P,2011ApJ...736....2O,2014ApJ...796..106J,2019MNRAS.484..687M}. This geometrical configuration is expected to result in beaming of the hard X-ray emissions from the funnel. Therefore, the beaming effect could potentially amplify high-energy variability and render the high-energy QPO waveform narrower in comparison to the low-energy one (see Figures~\ref{fig:2} and \ref{fig:3}). Additionally, there may be an accretion rate dependence of the beaming factor, expressed as $b\propto \dot{m}^{-2}$ \citep{2009MNRAS.393L..41K}, which could further enhance the amplitude of high-energy variability. Given that the high-energy variability exhibits a greater amplitude than the low-energy variability, we would expect the hardness ratio to be higher during the peak phases of the variability cycle. The tentative U-shaped feature in rms spectra of observations 101 and 201, however, remains inconclusive due to limited photon statistics; future advanced missions like \textit{eXTP} \citep{2019SCPMA..6229502Z} and \textit{Athena} \citep{2013arXiv1306.2307N} will be critical to confirm its physical origin.

The relatively longer exposure times of the three \emph{XMM-Newton} observations compared to previous observations allow us to robustly detect three periodic flux dips. These dips are likely recurrent, with a period of approximately 2 days, which is consistent with the orbital period of M51 ULX-7 \citep{2020ApJ...895...60R}. The estimated width of these dips are approximately $\sim0.15$--$0.4$ days, aligning with the dip features observed by \emph{Chandra} in 2012 \citep{2021ApJ...909....5H,2021ApJ...909...50V}. Given the similarities in the properties of the dip features observed by \emph{XMM-Newton} and \emph{Chandra}, we propose that they share the same origin, which is related to the binary orbital period. Since the total eclipse was not observed, \citet{2021ApJ...909....5H} suggested that the periodic dips result from obscuration of the emission from the accreting pulsar due to the vertical structure in the stream-disk interaction region or the atmosphere of the companion star. 

\begin{acknowledgments}
We are grateful to the anonymous referee for constructive comments that helped us improve this paper. This research has made use of data obtained from the High Energy Astrophysics Science Archive Research Center (HEASARC), provided by NASA’s Goddard Space Flight Center. This work is supported by the National Key R\&D Program of China (2021YFA0718500) and the National Natural Science Foundation of China (NSFC) under grants, 12025301 and 12333007. This work is partially supported by International Partnership Program of Chinese Academy of Sciences (Grant No.113111KYSB20190020). 
\end{acknowledgments}

%




\appendix
\section{Power Density Spectrum Analysis}
\label{appendix1}
The power density spectra (PDSs) of the three observations of M51 ULX-7 were constructed and modeled following the procedure described by \citet{2024A&A...689A.284I}. The 0.3–10 keV PDSs were computed with a bin time of 5 s in 1 segment, resulting in the minimum and Nyquist frequencies of $\sim6\times10^{-6}$ and 0.1 Hz, respectively. The PDSs were then logarithmically rebinned in frequency with a bin size increasing by a factor of 1.2 and normalized to units of fractional rms squared per Hz \citep{1990A&A...230..103B}. We fitted the PDSs with a model containing two Lorentzian components, a power-law component, and a constant component \citep[see Equation 1 of][]{2024A&A...689A.284I}. The two Lorentzians describe the broad shoulder at higher frequencies and the mHz QPO at lower frequencies, while the power-law component models the red noise dominant at frequencies $f<10^{-5}$ Hz. The constant component accounts for the white noise. Our fitting results are consistent with those presented in Table 2 of \citet{2024A&A...689A.284I}. Figure~\ref{fig:A} shows the PDS analysis of Observation 101 as an example, with the Poisson noise (derived from the best-fit constant component) subtracted.

\setcounter{figure}{0}
\renewcommand{\thefigure}{A\arabic{figure}}

\begin{figure}
\centering
    \includegraphics[width=0.5\textwidth]{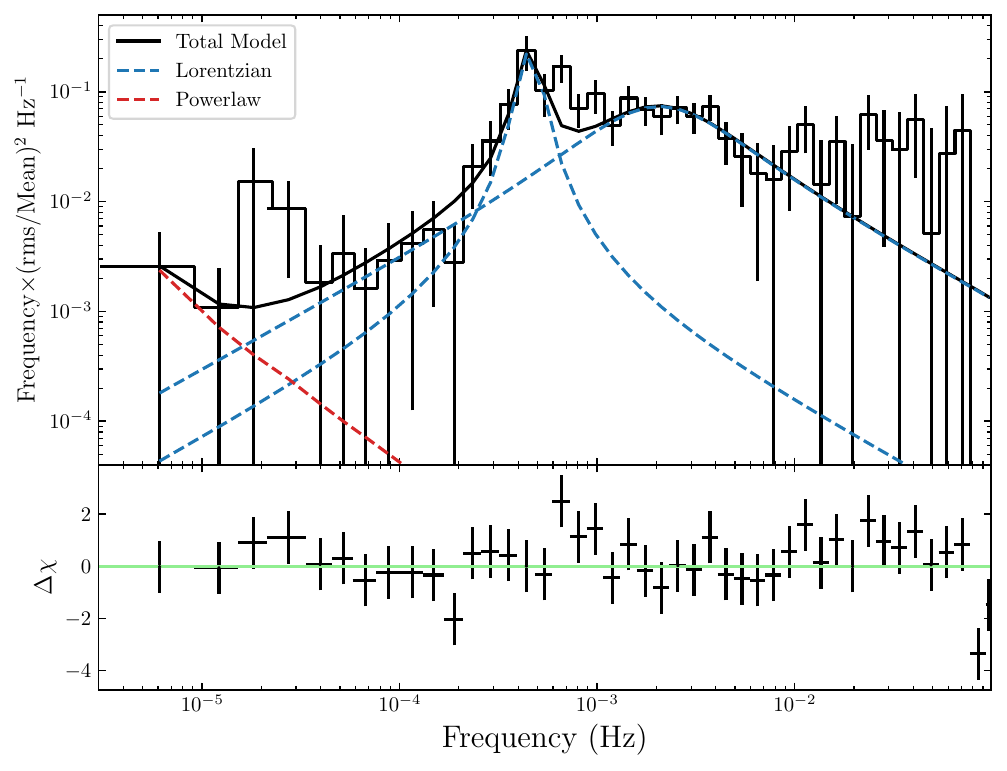}
    \caption{A PDS analysis of observation 101. Top panel: 0.3--10 keV Poisson-noise-subtracted PDS fitted with a model (solid black line) consisting of two Lorentzian (dashed blue lines) and a power-law (dashed red line) components. Bottom pabel: Residuals of the PDS with respect to the derived model.} \label{fig:A}
\end{figure}


\bibliography{sample631}{}
\bibliographystyle{aasjournal}



\end{document}